\documentstyle[12pt]{article} 
\input psfig.sty

\def\Title#1{\begin{center} {\Large #1 } \end{center}}
\def\Author#1{\begin{center}{ \sc #1} \end{center}}
\def\Address#1{\begin{center}{ \it #1} \end{center}}

\def\doeack{\footnote{Work supported by the Department of Energy,
                     contract DE--AC03--76SF00515.}}
\def\SLAC{Stanford Linear Accelerator Center\\
    Stanford University, Stanford, California 94309 USA}

\newenvironment{Abstract}{\begin{quotation} \begin{center}
                       ABSTRACT
     \end{center}\bigskip  }{\end{quotation}}
\def\beq{\begin{equation}}
\def\eeq#1{\label{#1}\end{equation}}
\def\eeqn{\end{equation}}
\def\beqa{\begin{eqnarray}}
\def\eeqa#1{\label{#1}\end{eqnarray}}
\def\eeqan{\end{eqnarray}}

\def\Re{{\cal R \mskip-4mu \lower.1ex \hbox{\it e}\,}}
\def\Im{{\cal I \mskip-5mu \lower.1ex \hbox{\it m}\,}}
\def\nn{\noindent}
\def\ie{{\it i.e.}}
\def\eg{{\it e.g.}}

\def\etal{{\it et al.}}

\def\sub#1{_{\lower.25ex\hbox{$\scriptstyle#1$}}}
\def\sul#1{_{\kern-.1em#1}}
\def\sll#1{_{\kern-.2em#1}}  
\def\sbl#1{_{\kern-.1em\lower.25ex\hbox{$\scriptstyle#1$}}}
\def\ssb#1{_{\lower.25ex\hbox{$\scriptscriptstyle#1$}}}
\def\sbb#1{_{\lower.4ex\hbox{$\scriptstyle#1$}}}

\def\to{\rightarrow}
\def\mh{\ifmmode m\sbl H \else $m\sbl H$\fi}
\def\mch{\ifmmode m_{H^\pm} \else $m_{H^\pm}$\fi}
\def\mt{\ifmmode m_t\else $m_t$\fi}
\def\mc{\ifmmode m_c\else $m_c$\fi}
\def\mz{\ifmmode M_Z\else $M_Z$\fi}
\def\mw{\ifmmode M_W\else $M_W$\fi}
\def\mws{\ifmmode M_W^2 \else $M_W^2$\fi}
\def\mhs{\ifmmode m_H^2 \else $m_H^2$\fi}   
\def\mzs{\ifmmode M_Z^2 \else $M_Z^2$\fi}
\def\mts{\ifmmode m_t^2 \else $m_t^2$\fi}
\def\mcs{\ifmmode m_c^2 \else $m_c^2$\fi}
\def\mchs{\ifmmode m_{H^\pm}^2 \else $m_{H^\pm}^2$\fi}
\def\ztwo{\ifmmode Z_2\else $Z_2$\fi}
\def\zone{\ifmmode Z_1\else $Z_1$\fi}
\def\mtwo{\ifmmode M_2\else $M_2$\fi}
\def\mone{\ifmmode M_1\else $M_1$\fi}
\def\tb{\ifmmode \tan\beta \else $\tan\beta$\fi}
\def\xw{\ifmmode x\sub w\else $x\sub w$\fi}
\def\ch{\ifmmode H^\pm \else $H^\pm$\fi}
\def\lum{\ifmmode {\cal L}\else ${\cal L}$\fi}
\def\inpb{\ifmmode {\rm pb}^{-1}\else ${\rm pb}^{-1}$\fi}
\def\infb{\ifmmode {\rm fb}^{-1}\else ${\rm fb}^{-1}$\fi}
\def\epem{\ifmmode e^+e^-\else $e^+e^-$\fi}
\def\ppb{\ifmmode \bar pp\else $\bar pp$\fi}

\def\bsg{\ifmmode b\rightarrow s\gamma \else $b\rightarrow s\gamma$\fi}
 
\newskip\zatskip \zatskip=0pt plus0pt minus0pt
\def\matth{\mathsurround=0pt}

\def\atversim#1#2{\lower0.7ex\vbox{\baselineskip\zatskip\lineskip\zatskip
  \lineskiplimit 0pt\ialign{$\matth#1\hfil##\hfil$\crcr#2\crcr\sim\crcr}}}

\begin{document}
\rightline{\vbox{\halign{&#\hfil\cr
&SLAC-PUB-7284\cr
&September 1996\cr}}}
\vspace{0.8in} 
\Title{Searches for Scalar and Vector Leptoquarks at Future Hadron Colliders}
\bigskip
\Author{Thomas G. Rizzo\doeack}
\Address{\SLAC}
\bigskip
\begin{Abstract}
The search reaches for both scalar($S$) and vector($V$) leptoquarks at future 
hadron colliders are summarized. In particular we evaluate the production 
cross sections of both leptoquark types at TeV33 and LHC as well as the 
proposed 60 and 200 TeV colliders through both quark-antiquark annihilation 
and gluon-gluon fusion: $q \bar q, gg \to SS,VV$. Experiments at these 
machines should easily discover such particles if their masses are not in 
excess of the few TeV range. 
\end{Abstract}
\bigskip
\vskip1.0in
\begin{center}
To appear in the {\it Proceedings of the 1996 DPF/DPB Summer Study on New
 Directions for High Energy Physics-Snowmass96}, Snowmass, CO, 
25 June-12 July, 1996. 
\end{center}
%
\bigskip
\def\thefootnote{\fnsymbol{footnote}}
\setcounter{footnote}{0}
\newpage

\section{Introduction}
Many extensions of the standard model(SM) which place quarks and leptons on
a symmetric footing predict the existence of leptoquarks, which are spin-0 
or 1, color-triplet objects that couple to a $q\ell$ or $\bar q\ell$ 
pair{\cite {bigref}}. While 
these particles may be sought indirectly through their influence on low energy 
processes{\cite {sacha}}, the most promising approach is via direct production 
at colliders. In particular, searches for leptoquarks at 
LEP{\cite {lep}}, HERA{\cite {hera}}, and the Tevatron{\cite {tev}} 
have already been performed, in most cases concentrating on the specific 
scenario of scalar leptoquarks. Based on both the direct and indirect 
searches we might expect that 
if leptoquarks exist their masses must be above a few hundred GeV, and 
possibly up in the TeV range. In this paper we 
will examine the search reach for both scalar and vector leptoquarks at future 
hadron colliders. The production rates for leptoquarks at such colliders will 
be shown to be sufficiently large so that particles of this type in the TeV 
mass range and above become accessible. In addition, we will see that the 
size of the production cross section alone is 
sufficient to distinguish scalar from vector leptoquark types.

\section{Leptoquark Pair Production}
Leptoquarks can be produced either singly or in pairs in hadronic 
collisions. The cross section for single production, however, relies on the 
size of the {\it a priori} unknown Yukawa couplings of the leptoquark and is 
therefore model dependent. Pair production, on the otherhand, proceeds 
through QCD interactions and thus depends {\it only} on 
the leptoquark spin and the fact that it is a color triplet field. It has been 
shown in Ref.{\cite {scalar}} that this 
mechanism will be dominant unless the Yukawa couplings, which are governed by 
the electroweak interactions, are rather large, \ie, of order electromagnetic 
strength or greater. This is an important result in that 
the production of both scalar and vector leptoquarks at hadron colliders is 
not dependent upon the electroweak properties of these particles. Of course, 
the converse is also true, \ie, the production properties cannot be used to 
probe the detailed nature of the leptoquark type. 

The production cross section for pairs of scalar leptoquarks($S$) in either 
$gg$ or $q\bar q$ 
collisions is easily calculated and has been available for some time; we use 
the results of Hewett \etal ~in 
Ref.{\cite {scalar}} in what follows. Given the mass of $S$ there are no real 
ambiguities in this calculation except for the possible inclusion of a 
$K-$factor (which we omit) to account for higher order QCD corrections and 
leads to a slight enhancement in the rate.

\vspace*{-0.5cm}
\nn
\begin{figure}[htbp]
\centerline{
\psfig{figure=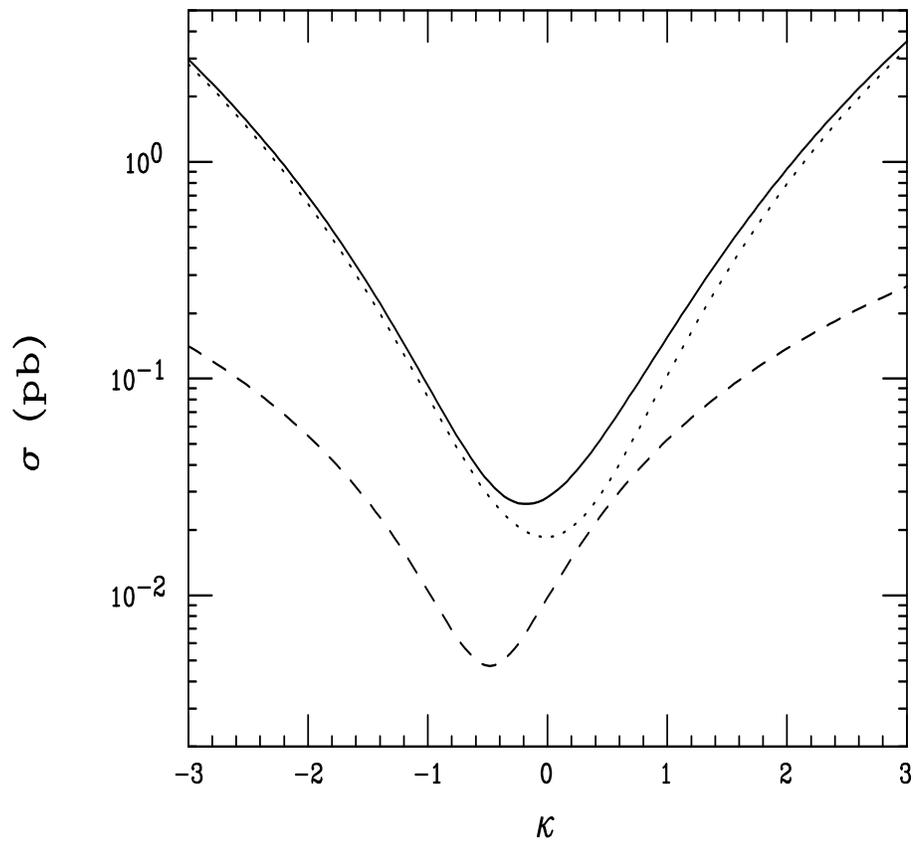,height=14cm,width=14cm,angle=-90}}
\vspace*{-0.6cm}
\caption{Production cross section for a pair of 1 TeV vector leptoquarks at 
the LHC as a function of $\kappa$. The dotted(dashed) curve 
corresponds to the $gg$($q\bar q$) production subprocess whereas the solid 
curve is their sum.}
\label{compare}
\end{figure}
\vspace*{0.1mm}

The vector leptoquark($V$) case is not as straightforward. 
In order to calculate the $gg \to VV$ cross section we need to determine both 
the trilinear $gVV$ and quartic $ggVV$ couplings, which may naively at first 
glance appear to be unknown. (For the $q \bar q$ subprocess, only the $gVV$ 
coupling is required.) However, in any realistic model wherein vector 
leptoquarks appear and are {\it fundamental} objects, they will be 
the gauge bosons of an extended gauge group like $SU(5)$. In this case 
the $gVV$ and 
$ggVV$ couplings are {\it completely} fixed by extended gauge invariance.  
These particular couplings will also insure that the subprocess cross sections  
obey tree-level unitarity, as is the hallmark of all gauge theories. 
Of course, it might be that the appearance of vector 
leptoquarks is simply some low energy manifestation of a 
more fundamental theory at a higher scale and that these particles may even 
be composite. In this case so-called `anomalous' couplings in both the $gVV$ 
and $ggVV$ vertices can appear. One possible coupling of this type 
is an `anomalous 
chromomagnetic moment', usually described in the literature by the parameter 
$\kappa$, which takes the value of unity in the more realistic gauge theory 
case. Among these `anomalous couplings', the term which induces $\kappa$ is
quite special in that it is the only one that conserves $CP$ and is of
dimension 4.

\vspace*{-0.5cm}
\nn
\begin{figure}[htbp]
\centerline{
\psfig{figure=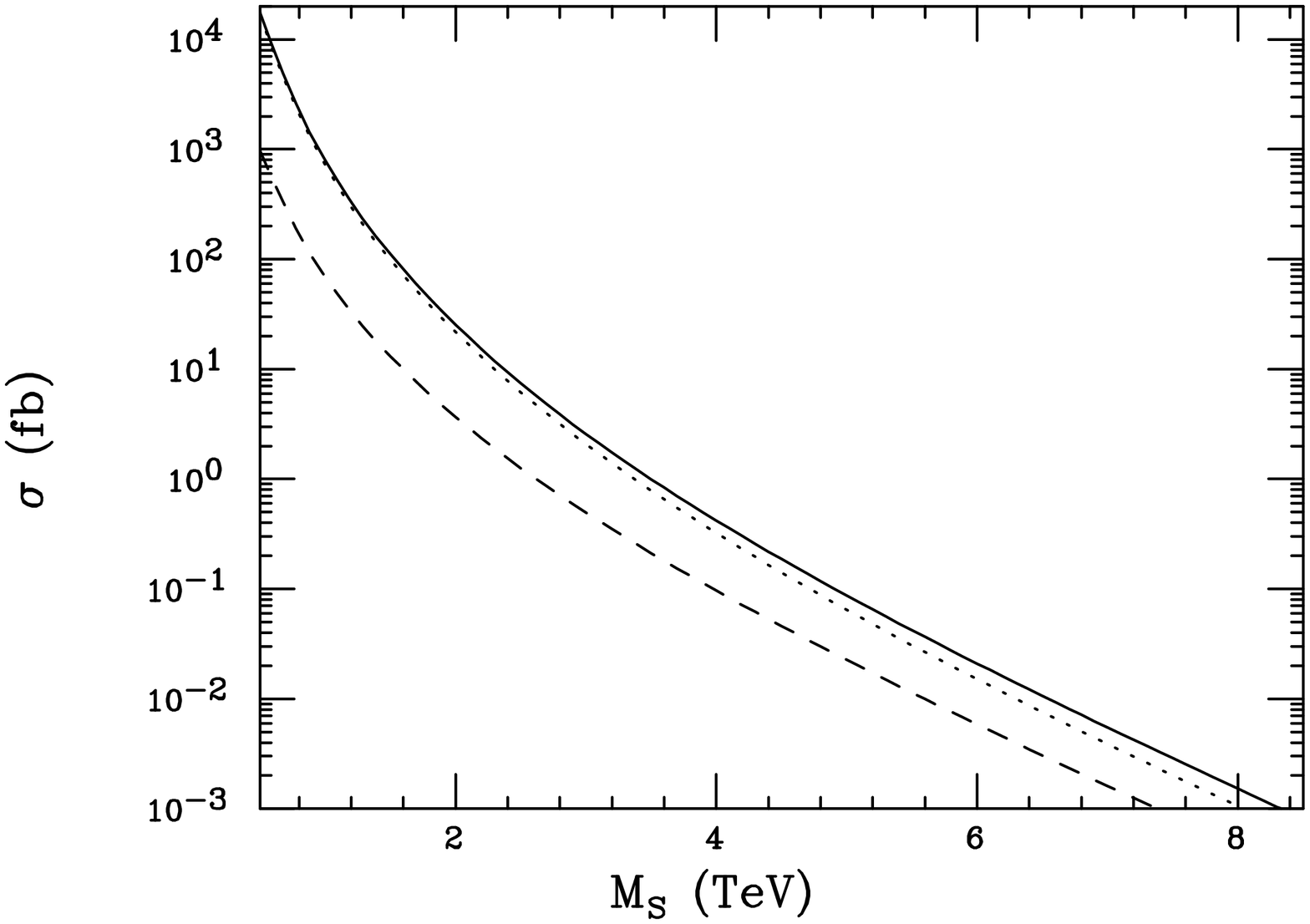,height=9.1cm,width=9.1cm,angle=-90}
\hspace*{-5mm}
\psfig{figure=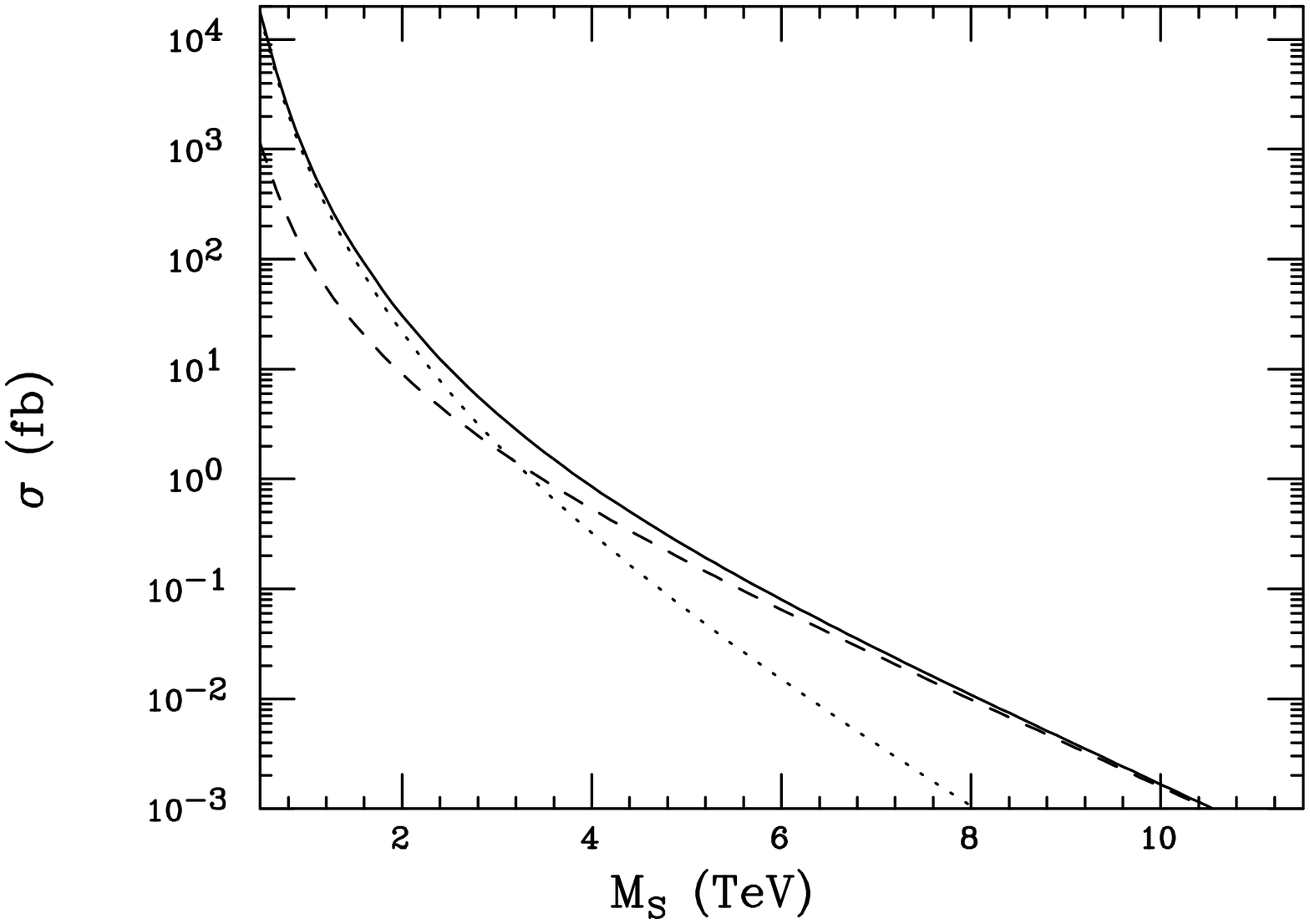,height=9.1cm,width=9.1cm,angle=-90}}
\vspace*{-0.6cm}
\caption{Scalar leptoquark pair production cross section as a function of mass 
at a 60 TeV $pp$(left) or $p\bar p$(right) LSGNA collider. The 
dotted(dashed) curve 
corresponds to the $gg$($q\bar q$) production subprocess whereas the solid 
curve is their sum. MRSA$'$ parton densities are employed.}
\label{s60}
\end{figure}
\vspace*{0.1mm}

The Feynman rules for the vector leptoquark-gluon interactions
can then be derived from the following effective Lagrangian which includes the
most general set of $SU(3)_c$ gauge invariant, $CP$-conserving operators of 
dimension 4 (or less)
\begin{equation}
{\cal L}_V =-{1\over 2} F^\dagger_{\mu\nu}F^{\mu\nu}+M_V^2V^\dagger_\mu V^\mu
-ig_s\kappa V^\dagger_\mu G^{\mu\nu}V_\nu  \,.
\end{equation}
Here, $G_{\mu\nu}$ is the usual gluon field strength tensor,
$V_\mu$ is the vector leptoquark field and $F_{\mu\nu}=D_\mu V_\nu-D_\nu
V_\mu$, where $D_\mu=\partial_\mu+ig_sT^a G^a_\mu$ is the gauge
covariant derivative (with respect to $SU(3)$ color), $G^a_\mu$ is the
gluon field and the $SU(3)$ generator $T^a$ is taken in the triplet
representation.  In most of the numerical results that follow we 
will assume $\kappa=1$, \ie, we will 
assume that $V$ is indeed a gauge boson and use the results of Hewett \etal ~in 
Ref.{\cite {vlq}} for the evaluation of production rates. The cross sections 
for other nearby values of $\kappa$ are generally qualitatively comparable 
as is demonstrated by the results shown in Fig.~\ref{compare} for the case of 
the pair production of 1 TeV vector leptoquarks at the LHC. If the vector 
leptoquark is {\it not} a gauge boson then we might, \eg, expect it to be 
minimally coupled to the gluon field, as discussed by Bl\"umlein and 
R\"uckl{\cite {vlq}}. In this case we have instead that $\kappa=0$. The cross 
section in this case, as can be seen from the figure, is somewhat smaller than 
in the situation where $V$ is a vector boson with $\kappa=1$. We remind the 
reader 
that changes in $\kappa$ will also lead to modifications in the distributions 
associated with vector leptoquark pair production but these are subjects are 
beyond the scope of the present analysis and will be discussed elsewhere.

\vspace*{-0.5cm}
\nn
\begin{figure}[htbp]
\centerline{
\psfig{figure=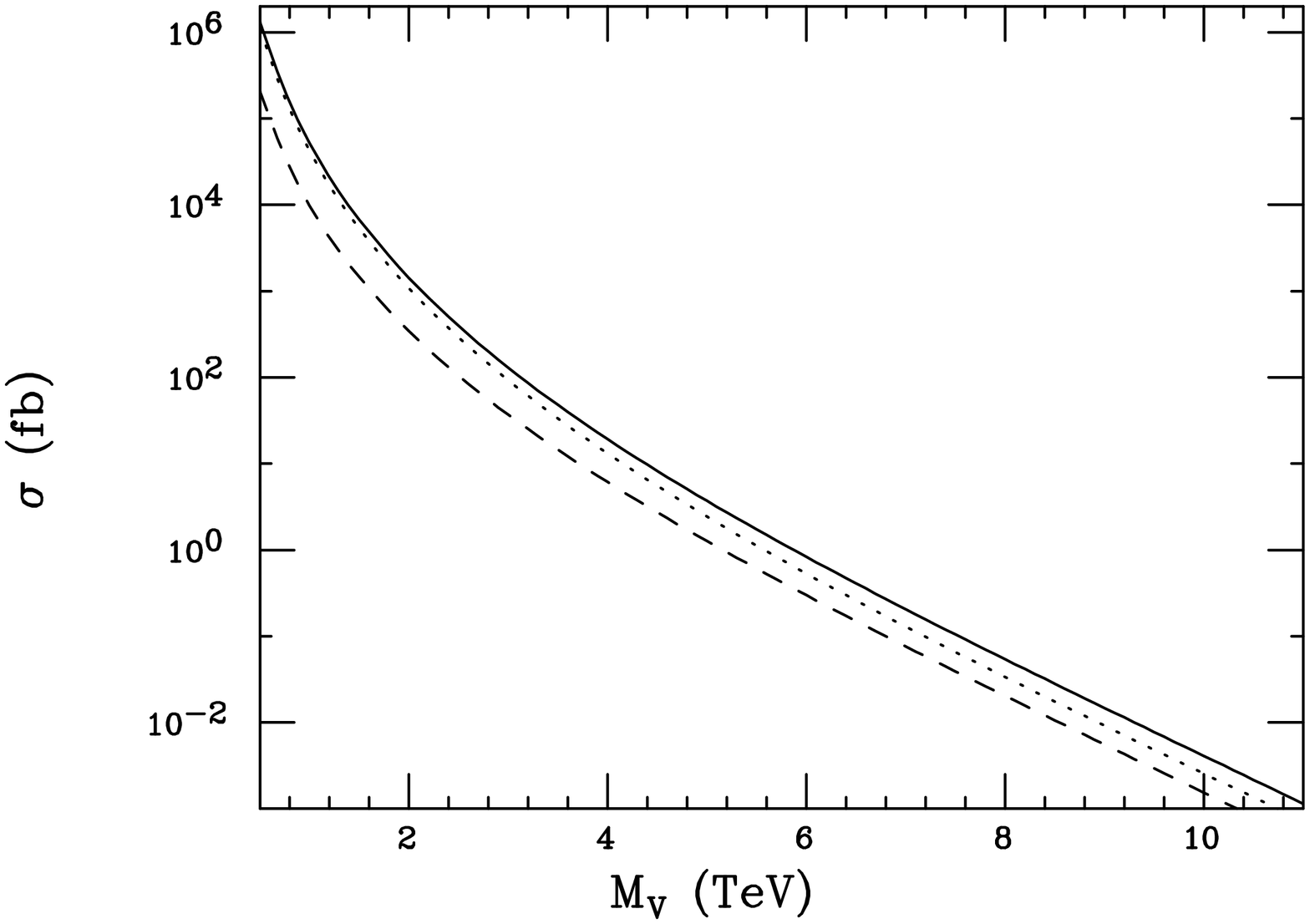,height=9.1cm,width=9.1cm,angle=-90}
\hspace*{-5mm}
\psfig{figure=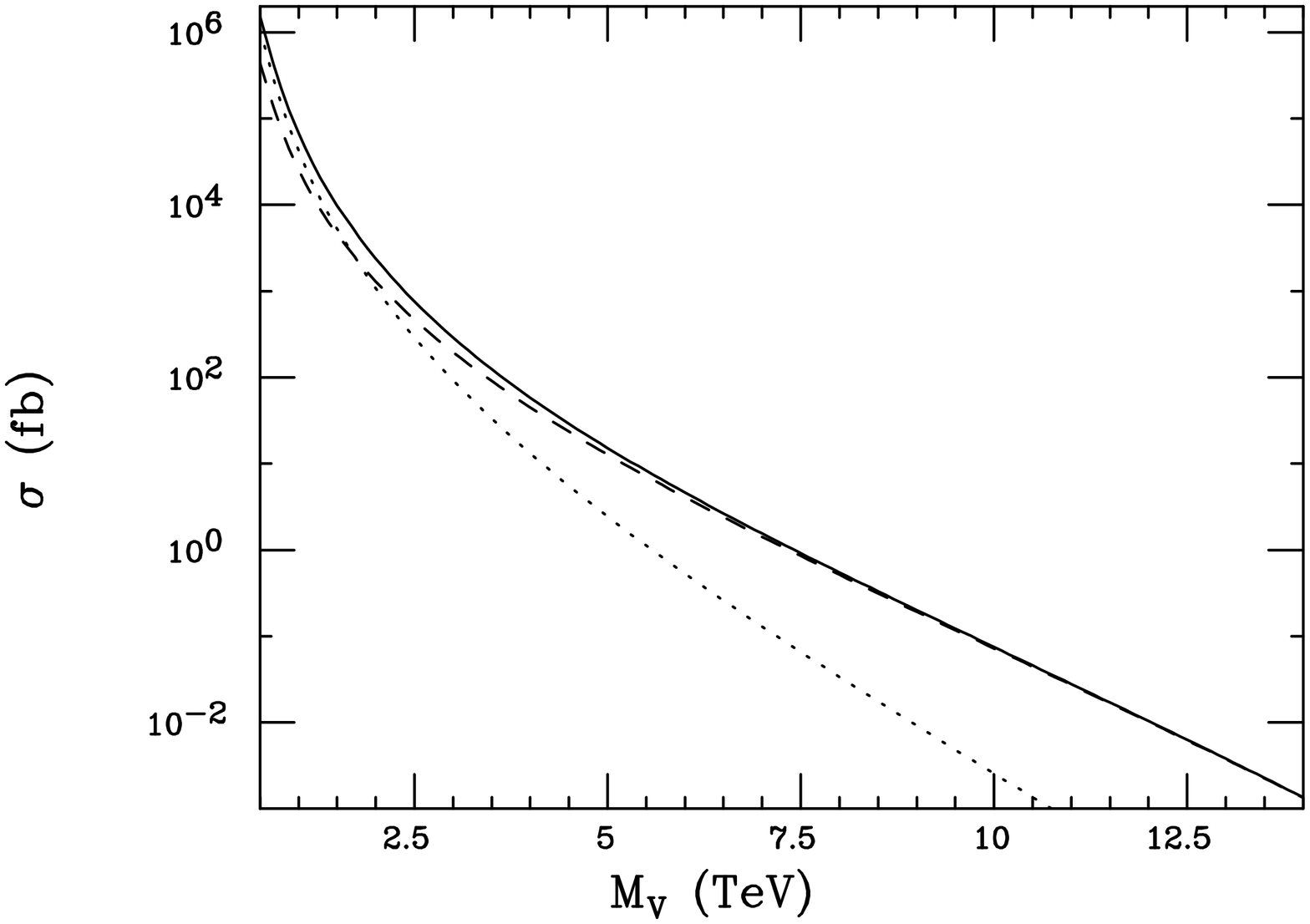,height=9.1cm,width=9.1cm,angle=-90}}
\vspace*{-0.6cm}
\caption{Same as the previous figure but now for a spin-1 vector leptoquark 
with $\kappa=1$. }
\label{v60}
\end{figure}
\vspace*{0.1mm}

\section{Results}

\begin{figure}[htbp]
\centerline{
\psfig{figure=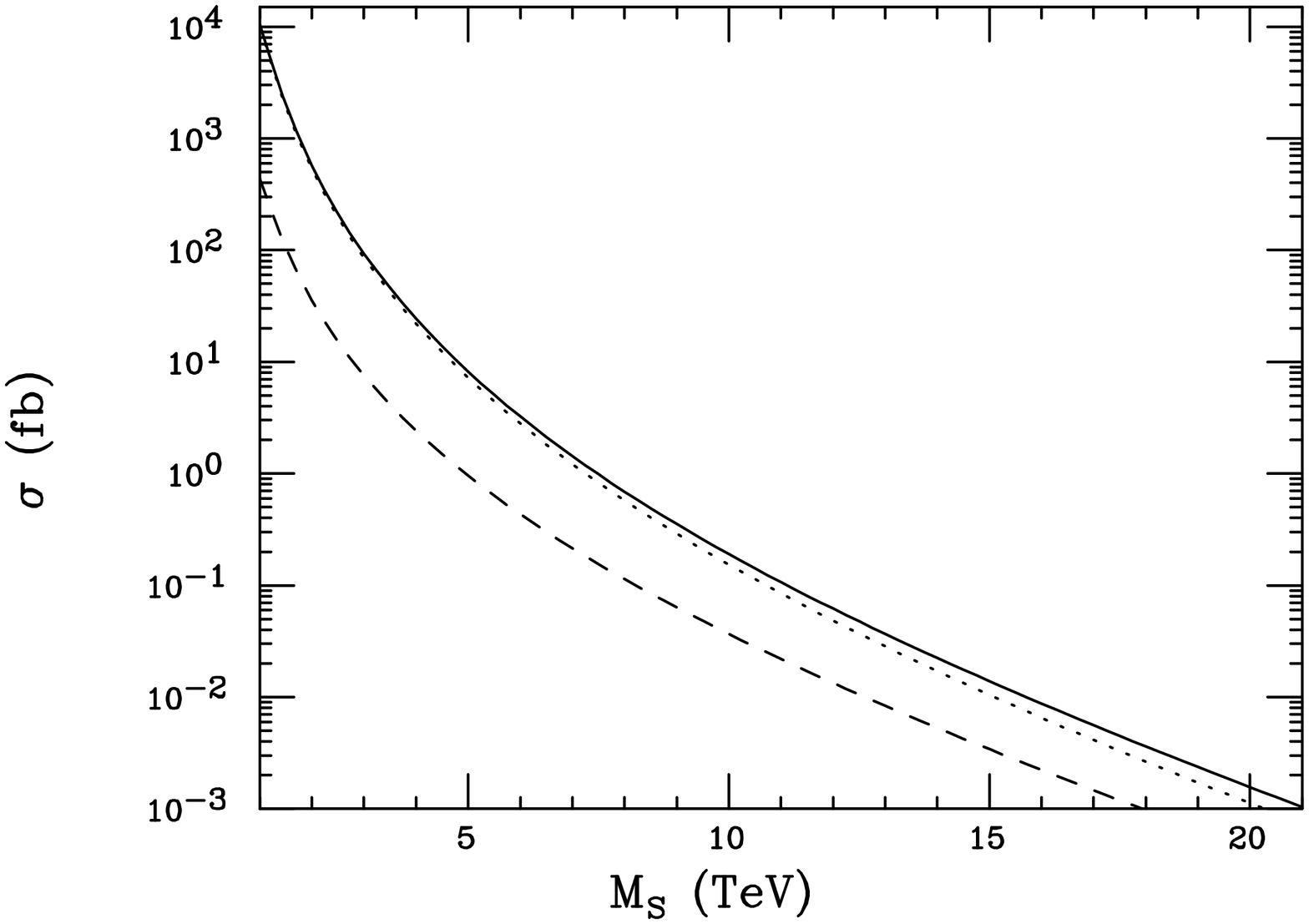,height=9.1cm,width=9.1cm,angle=-90}
\hspace*{-5mm}
\psfig{figure=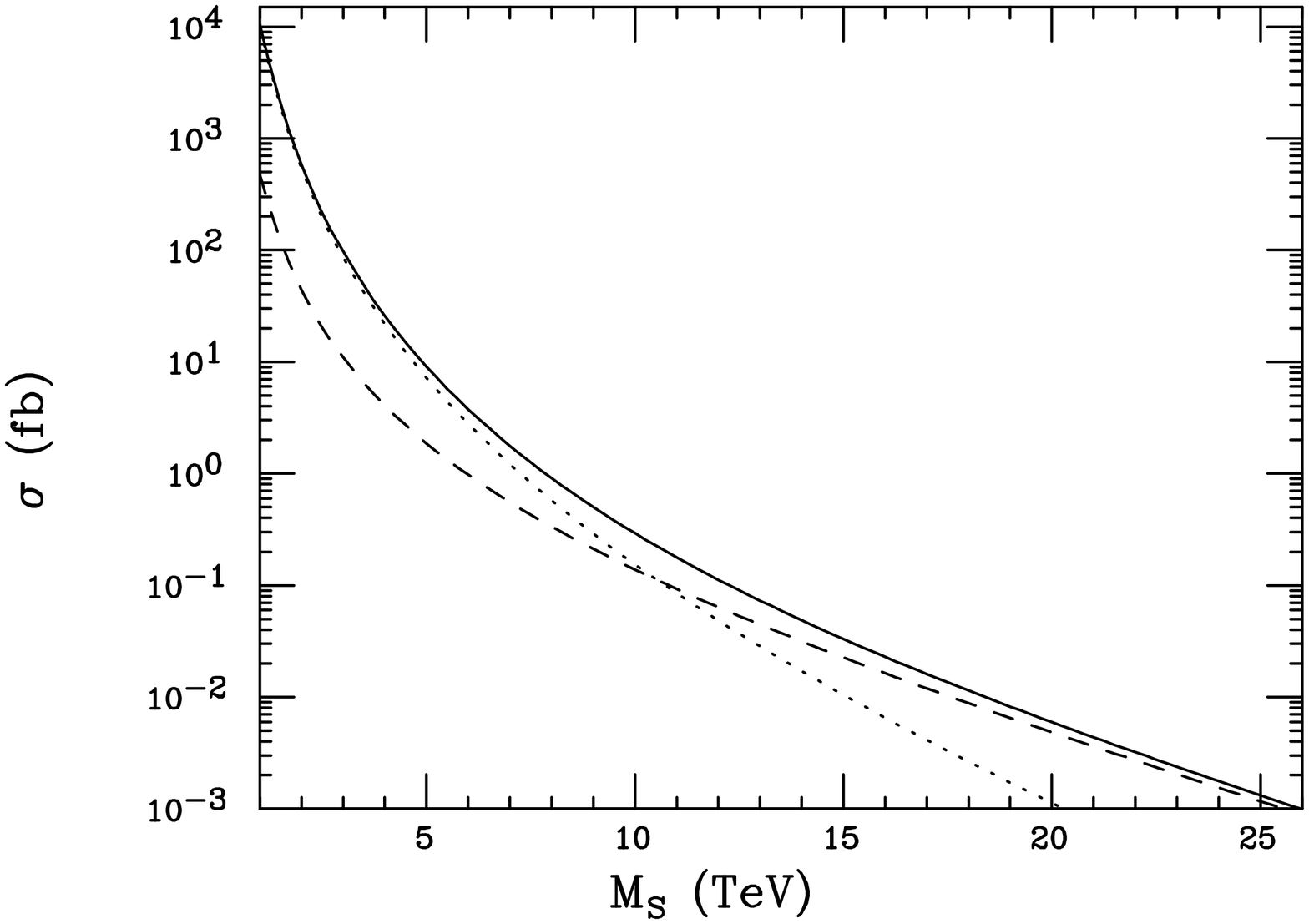,height=9.1cm,width=9.1cm,angle=-90}}
\vspace*{-0.6cm}
\caption{Same as Fig.2 but now at the 200 TeV PIPETRON collider.}
\label{s200}
\end{figure}

\begin{figure}[htbp]
\centerline{
\psfig{figure=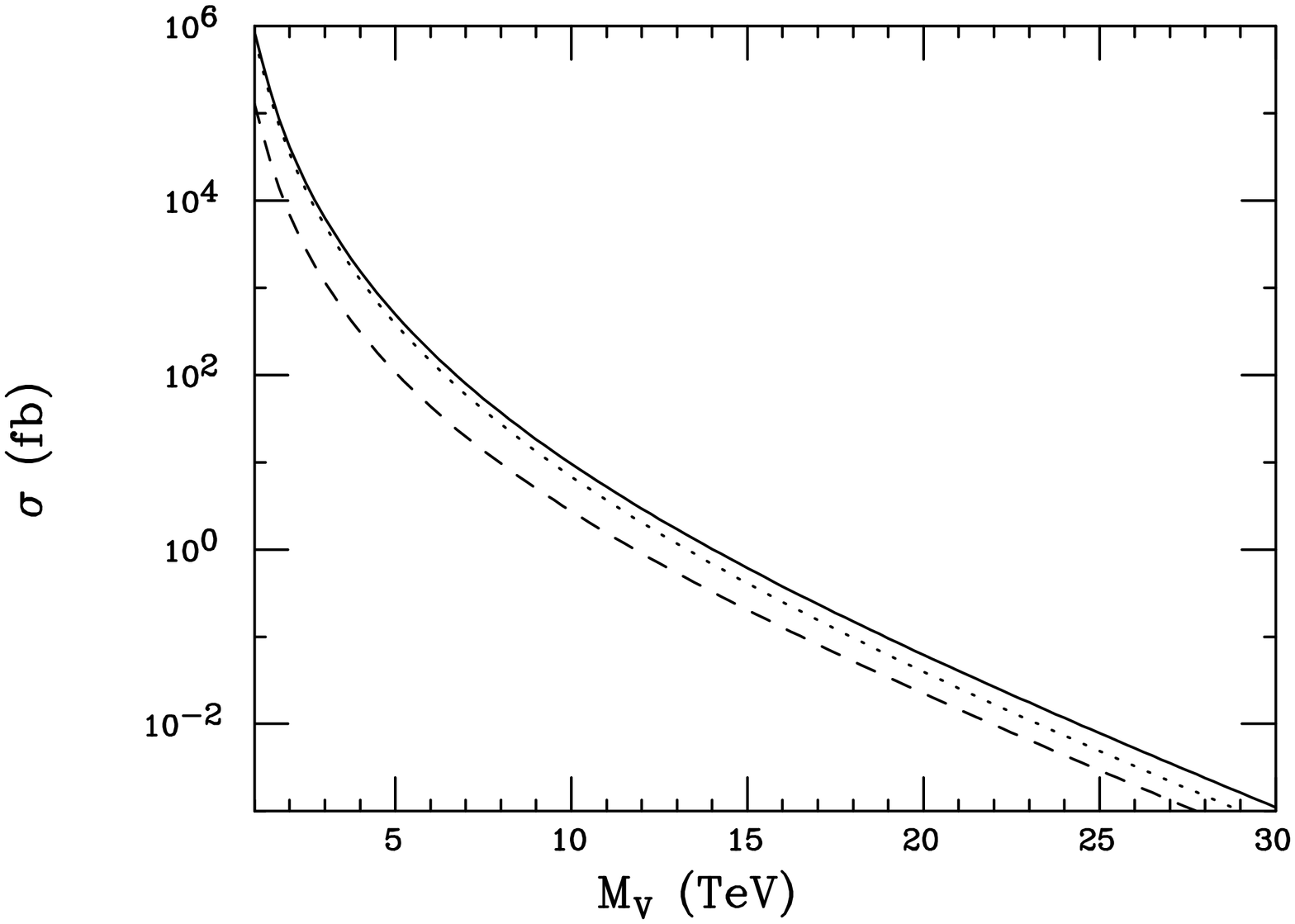,height=9.1cm,width=9.1cm,angle=-90}
\hspace*{-5mm}
\psfig{figure=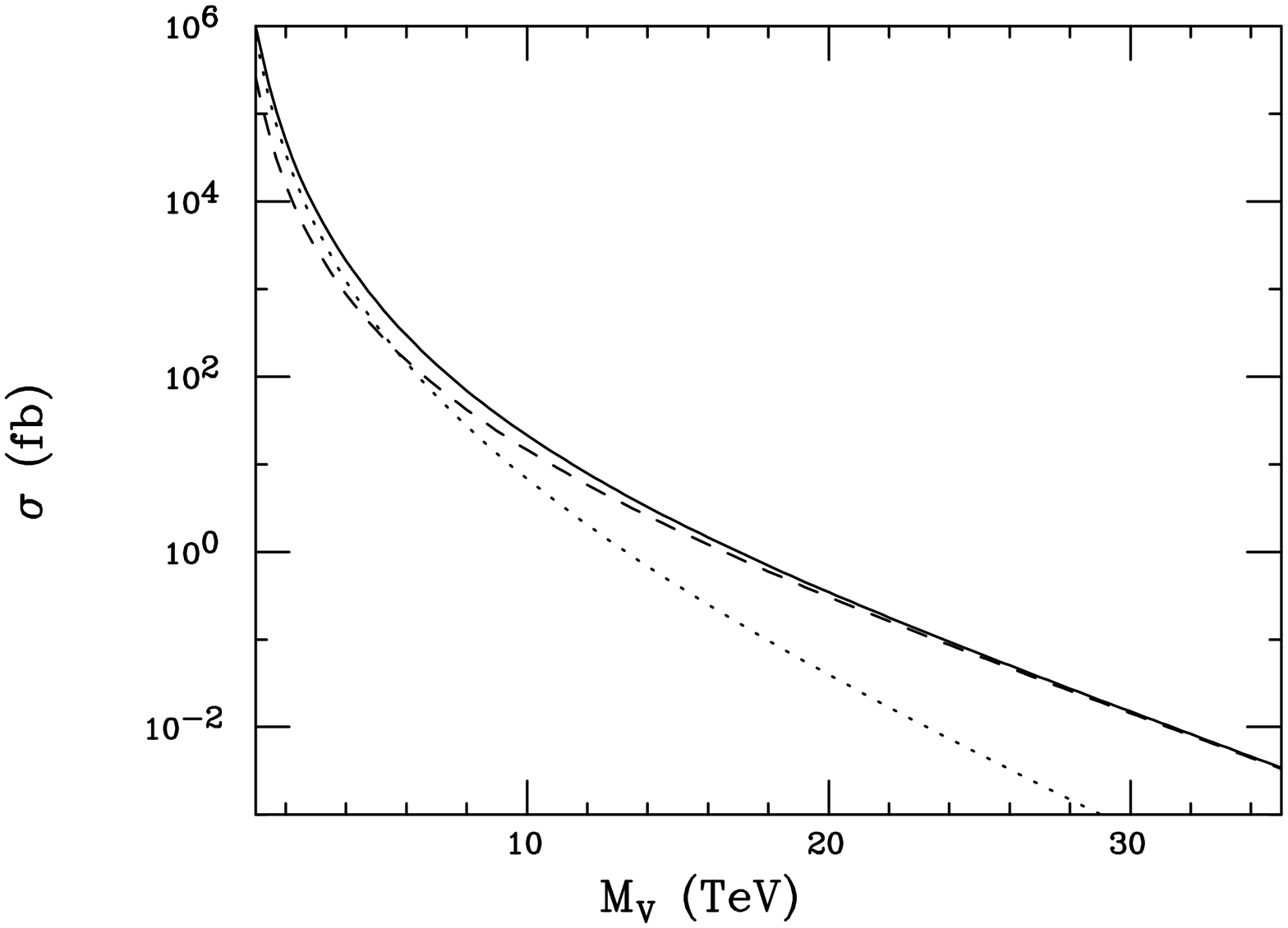,height=9.1cm,width=9.1cm,angle=-90}}
\vspace*{-0.6cm}
\caption{Same as Fig.3 but now for the 200 TeV PIPETRON collider.}
\label{v200}
\end{figure}

We now turn to some numerical results. We will consider the production of only 
a single type of leptoquark at a time and ignore the possibility of a 
degenerate multiplet of leptoquarks being produced simultaneously. Here 
we concentrate on cross 
sections for $S$ and $V$ pair production at the $\sqrt s$=60 (LSGNA) and 
200 (PIPETRON) TeV machines, 
which are displayed in Figures~\ref{s60},~\ref{v60},~\ref{s200} and~\ref{v200}, 
since the corresponding results for the Tevatron and LHC can be found in, \eg,  
Ref.{\cite {dj}}. In these figures, the contributions of the two distinct 
subprocesses $gg\to SS,VV$ and $q\bar q \to SS,VV$ are separately 
displayed together with 
their sum. From Figures~\ref{s60} and~\ref{v60} several conclusions are 
immediately obvious for leptoquark production at the $\sqrt s=60$ TeV 
collider: ($i$) The vector leptoquark cross section is 
substantially larger than that for scalars in both $pp$ and $p\bar p$ 
collisions since the rates for both $gg\to VV$ and $q\bar q\to VV$ are larger 
than their scalar counterparts. ($ii$) Due to the contribution of the $q\bar q$ 
production mode, 
$p\bar p$ colliders have larger leptoquark cross sections than do $pp$ 
colliders. For example, the ratio of $p\bar p$ to $pp$ cross sections for a 
4(6) TeV scalar(vector) leptoquark is approximately 2(6) at $\sqrt s=60$ TeV. 
At $pp$ machines, for both vector and scalar leptoquarks, the cross 
sections are dominated by the $gg$ process out to the machine's anticipated 
mass reach. In the $\sqrt s=60$ TeV $p\bar p $ case, the $q\bar q$ process 
dominates over $gg$ for masses 
greater than about 3.0(1.8) TeV for scalar(vector) leptoquarks. The mass 
reaches for the 60 TeV machine can be found in Table I. 

\vspace*{-0.5cm}
\nn
\vspace*{0.1mm}

At $\sqrt s$=200 TeV, the patterns observed at 60 TeV are repeated. 
For example, the ratio of $p\bar p$ to $pp$ cross sections for a 
10(15) TeV scalar(vector) leptoquark is approximately 1.5(3.5). 
In the $p\bar p $ collider mode, the $q\bar q$ process dominates over $gg$ 
for masses greater than about 10(6) TeV for scalar(vector) leptoquarks.

\vspace*{-0.5cm}
\nn
\vspace*{0.1mm}

Table~\ref{leptos} summarizes and compares the search reaches for both scalar 
and vector leptoquarks at the Tevatron and LHC as well as the hypothetical 
60 and 200 TeV $pp$ and $p\bar p$ colliders. Our results for the Tevatron 
confirm the expectations of the TeV2000 Study Group {\cite {tev2000}}, who also 
assume the 10 event discovery limit,  while those obtained for 
the LHC are somewhat smaller{\cite {wrochna}} than that given by a fast 
CMS detector simulation. As discussed above, the larger 
cross sections for vector leptoquarks results in higher search reaches at all 
machines. Similarly, the larger $q\bar q$ subprocess contribution to the total 
cross section at $p\bar p$ machines leads to a greater reach for both scalar 
and vector leptoquarks in this collision mode. 

It is clear from this table 
that future hadron colliders will be able to significantly extend the present 
search reaches for scalar and vector leptoquarks.

\begin{table}[htbp]
\centering
\begin{tabular}{lccc}
\hline
\hline
Machine  & ${\cal L}(fb^{-1})$  & $S$ & $V$ \\
\hline
LHC                    &  100  & 1.34(1.27)  & 2.1(2.0)    \\
60 TeV($pp$)           &  100  & 4.9(4.4)    & 7.6(7.0)    \\
60 TeV($p\bar p$)      &  100  & 5.7(5.2)    & 9.6(9.0)    \\
200 TeV($pp$)          &  1000 & 15.4(14.1)  & 24.2(23.3)  \\
200 TeV($p\bar p$)     &  1000 & 18.1(16.2)  & 31.1(29.0)  \\
TeV33                  &  30 & $\simeq 0.35$ & $\simeq 0.58$\\
\hline
\hline
\end{tabular}
\caption{Search reaches in TeV for scalar($S$) and vector($V$) leptoquarks at 
future hadron 
colliders assuming a branching fraction into a charged lepton plus a jet of 
unity($1/2$). For vector leptoquarks, $\kappa=1$ has been assumed and in both 
cases the MRSA$'$ parton densities have been employed. These results are 
based on the assumption of 10 signal events.}
\label{leptos}
\end{table}

\section{Acknowledgements}

The author would like to thanks S. Godfrey, J. Hewett, G. Wrochna and 
W. Merritt for discussions related to this work.

%
\def\MPL #1 #2 #3 {Mod.~Phys.~Lett.~{\bf#1},\ #2 (#3)}
\def\NPB #1 #2 #3 {Nucl.~Phys.~{\bf#1},\ #2 (#3)}
\def\PLB #1 #2 #3 {Phys.~Lett.~{\bf#1},\ #2 (#3)}
\def\PR #1 #2 #3 {Phys.~Rep.~{\bf#1},\ #2 (#3)}
\def\PRD #1 #2 #3 {Phys.~Rev.~{\bf#1},\ #2 (#3)}
\def\PRL #1 #2 #3 {Phys.~Rev.~Lett.~{\bf#1},\ #2 (#3)}
\def\RMP #1 #2 #3 {Rev.~Mod.~Phys.~{\bf#1},\ #2 (#3)}
\def\ZP #1 #2 #3 {Z.~Phys.~{\bf#1},\ #2 (#3)}
\def\IJMP #1 #2 #3 {Int.~J.~Mod.~Phys.~{\bf#1},\ #2 (#3)}

\end{document}